\documentclass[12pt]{article}
\usepackage{amsmath,amsfonts, amssymb, braket, bbold}
\usepackage{graphicx,multicol,multirow}
\usepackage{hyperref}
\usepackage{xcolor}
\usepackage{textpos, wrapfig}
\usepackage{titlesec}
\titleformat{\section}
 {\normalfont\fontfamily{put}\fontsize{12pt}{16pt}\bfseries\color{black}}
{\thesection}{1em}{}
\titleformat{\subsection}
 {\normalfont\fontfamily{put}\fontsize{12pt}{16pt}\bfseries\color{black}}
{\thesubsection}{1em}{}
\definecolor{maroon}{rgb}{.45, .2, .05}
\definecolor{medblue}{rgb}{.35, .4, 1}
\definecolor{darkblue}{rgb}{.2, .1, .65}
\definecolor{darkgreen}{rgb}{0, .7, 0}
\definecolor{lightgrey}{rgb}{.85, .90, .9}
\definecolor{brown}{rgb}{.25, .25, .25}
\definecolor{lightblue}{rgb}{.73, .83, .99}
\definecolor{lightbrown}{rgb}{.88, .8, .55}
\definecolor{goldenrod}{rgb}{.80392, .60784, .11373}
\definecolor{darkgoldenrod}{rgb}{.5451, .39608, .03137}
\definecolor{darkolivegreen}{rgb}{.33333, .41961, .18431}
\definecolor{darkred}{rgb}{.75, .15, .15}
\definecolor{cyellow}{rgb}{1.0, 0.812, 0.004} 
\definecolor{cred}{rgb}{1.0, 0.220, 0.224}
\definecolor{mellow}{rgb}{.847, .72, .525}
\definecolor{orange}{rgb}{1.00, 0.65, 0.00}
\definecolor{deepred}{rgb}{.90, .10, .10}
\definecolor{lemonchiffon}{rgb}{1.00, .98, .80}
\def \beq  {\begin{equation}}
\def \eeq  {\end{equation}}
\def \beqar {\begin{eqnarray}}
\def \eeqar {\end{eqnarray}}
\allowdisplaybreaks
\def\sqr#1#2{{\vcenter{\vbox{\hrule height.#2pt
\hbox{\vrule width.#2pt height#1pt \kern#1pt
\vrule width.#2pt}\hrule height.#2pt}}}}

\def\L {{\cal L}}

\def\vf {{\varphi}}

\def\Tr {{\rm Tr}}

\def\del {\partial}

\def\A {{\cal A}}

\def\C {{\cal C}}
\def\D {{\cal D}}

\def\H {{\cal H}}

\def \L {{\cal L}}

\def\vf {{\varphi}}

\def\half{\textstyle{1\over 2}}

\mathchardef\mhyphen="2D
\begin{document}
\fontfamily{bch}\fontsize{12pt}{17pt}\selectfont
\def \CMP {{Commun. Math. Phys.}}
\def \PRL {{Phys. Rev. Lett.}}
\def \PL {{Phys. Lett.}}
\def \NPBProc {{Nucl. Phys. B (Proc. Suppl.)}}
\def \NP {{Nucl. Phys.}}
\def \RMP {{Rev. Mod. Phys.}}
\def \JGP {{J. Geom. Phys.}}
\def \CQG {{Class. Quant. Grav.}}
\def \MPL {{Mod. Phys. Lett.}}
\def \IJMP {{ Int. J. Mod. Phys.}}
\def \JHEP {{JHEP}}
\def \PR {{Phys. Rev.}}
\def \JMP {{J. Math. Phys.}}
\def \GRG{{Gen. Rel. Grav.}}
\begin{titlepage}
\null\vspace{-62pt} \pagestyle{empty}
\begin{center}
\vspace{1.3truein} {\large\bfseries
~}
\\
{\large\bfseries A.P. Balachandran: Living in Physics}\\
\vskip .15in
{\bfseries (January 15, 1938-April 18, 2025)}\\
\vskip .5in
{\sc V.P. Nair}\\
\vskip .1in
{\sl Physics Department,
City College of New York, CUNY\\
New York, NY 10031}\\
 \vskip .1in
\begin{tabular}{r l}
{\sl E-mail}:&\!\!\!{\fontfamily{cmtt}\fontsize{11pt}{15pt}\selectfont 
 vpnair@ccny.cuny.edu}\\
\end{tabular}
\vskip .5in

\centerline{\large\bf ~}
\end{center}
A.P. Balachandran was a theoretical physicist of extraordinary
versatility and originality. 
In a career spanning over six decades, he made many
significant contributions to quantum field theory and
mathematical physics.
His influence extends beyond the impact of his papers, with
a large number of research students and collaborators.
Here, in an attempt at a coherent scientific biography,
 I try to contextualize and evaluate his many contributions.
 
\centerline{(To be published in the International Journal of Modern Physics A)}

\end{titlepage}
\pagestyle{plain} \setcounter{page}{2}

If asked about why we do physics, many answers can be given, from lofty noble aims of trying to understand the workings of the universe to the mundane one of 
having a career. But the fact of the matter is that many of us physicists 
realize that there is yet another impetus, unknown and perhaps unknowable:
we are just driven to do physics, quite simply we cannot do otherwise and
we willingly surrender to physics as it permeates our lives.

This was indeed certainly, and very visibly, true for
Bal who spent a lifetime in theoretical physics.
One could hold an erudite and enlightened
discussion with Bal, in English, Malayalam, Tamil,
Italian or Spanish, on politics, history, literature or movies, but 
every conversation would touch upon physics at some stage.

Aiyalam Parameswaran Balachandran, A.P. Balachandran, 
or Bal as he was known to his many students, collaborators 
and colleagues, was born in India, in the southern city of
Salem, in 1938.
He grew up in the coastal cities of
Ernakulam and Kozhikode and graduated
with his bachelor's degree
from the Madras Christian College in 1958.
He went on to obtain his Ph.D. from
the University of Madras in 1962.
He was the first doctoral graduate from the Institute of Mathematical
Sciences, which had been founded by his advisor Alladi Ramakrishnan.
He then spent a few months at the Theoretical Physics Institute
in Vienna, followed by two years at the Enrico Fermi Institute in Chicago.
He joined Syracuse University in 1964, eventually retiring as the Joel Dorman Steele
Professor of Physics in 2012.

A.P. Balachandran was a theoretical physicist of extraordinary versatility and originality.
His first papers in physics were from 1962, he wrote his last paper late in 2024, it was published in May 2025, a few days after he passed on.
(There is even some unfinished work with A. Pinzul, which may
appear shortly.)
 In a scientific lifetime spanning over six decades, he
wrote, or was a coauthor of, over 285 scientific articles and seven books.
He was a guide and mentor to more than 40 students and had
collaborators in every continent (except, Antarctica, I guess), 
of the order of several dozen.

In any appraisal or appreciation of Bal's contributions, it is important to
view them against the backdrop of developments in high energy and particle physics over a similar period of time.
The remarkable progress of theoretical physics in the last sixty years is manifest in high relief if we consider what we did not know in 1962.
Strong interactions were a complete unknown, despite interesting patterns revealed by the Eightfold Way. Quarks themselves were a concept of
the future. The $V$-$A$ theory had managed to pin down
low energy weak interactions, but the theory pointed to its own future obsolescence
by the breakdown of unitarity. Needless to say, there was no Standard Model,
nor was there string theory.
The Higgs mechanism was not known or at least not well-understood, despite Anderson's
work on superconductivity. Even the spontaneous breaking of global symmetries, including chiral symmetry breaking, was a nascent concept.
On the more theoretical side, perturbation theory for nonabelian gauge fields did not exist, there were no Faddeev-Popov ghosts and topology was viewed as something physicists did not need to know.
Topological solitons, monopoles, instantons were still
waiting for their debut, and anomalies, the example {\it par excellence}
of the importance of topology in physics, were yet to be discovered.
But by the end of the twentieth century, the essence, if not the microscopic details, of all these topics was understood and one could contemplate more daring questions such as a quantum theory of gravity, maybe using string theory, maybe via noncommutative geometry, maybe using a suitable algebra of observables or other concepts.

Bal's life in physics touched upon all of these topics, and while we cannot analyze and evaluate all of Bal's published works, here we will go over
some of his most original and impactful papers.

\section{Early work}

Going back to the 1960s, while the general understanding of strong interaction physics remained difficult, a combination of current algebra with the PCAC
(partial conservation of axial vector current)
hypothesis could be used for low energy scattering.
Weinberg had brilliantly summarized this in terms of an effective theory
(the nonlinear sigma model) \cite{weinberg}.
 Bal (with Gundzik and Nicodemi)
obtained a direct current-algebraic calculation of
$S$-wave and $P$-wave scattering lengths and effective range
for meson-baryon and meson--meson scattering \cite{gundzik}.
We may note that scattering lengths
 are a key point of comparison towards demonstrating
the equivalence between current algebra and the effective field theory description.
Scattering amplitudes remained the focus of Bal's work for many years,
with several papers (about 25) dealing with partial wave analysis, bounds and inequalities for them, as well as on the dual-resonance model, the early avatar of string theory.
Some of these papers were with early research
students W. Meggs and P. Ramond.
(Meggs moved on to research in medicine, Ramond went on to
work on many topics including grand unified theories, neutrino masses, etc.
and most famously discovered the
Ramond sector of string theory.)

\section {Monopoles, fiber bundles}

Monopoles dominated the next phase of Bal's research in physics.
Dirac had shown, already in 1931, how monopoles could be compatible with standard electrodynamics, if the product of electric and magnetic charges
was quantized; this is the so-called Dirac quantization condition \cite{dirac}. Attempts to produce an interacting field theory of charged particles and monopoles,
by Dirac himself \cite{dirac2}, and later by Schwinger \cite{schwinger1}, Zwanziger \cite{zwan} and others, had not produced a workable and easy-to-use quantum field theory.
But the key quantum jump in the development of these ideas came with the work of 't Hooft and Polyakov, who, independently, found monopole solutions 
in an $SU(2)$ gauge theory spontaneously broken to
$U(1)$ symmetry \cite{HP}. 
The unbroken $U(1)$ symmetry could be identified as the gauge symmetry of electromagnetism. 
The 't Hooft-Polyakov monopoles were nonperturbative solutions to the classical equations of motion. The developments in the functional approach to quantum field theory, relatively recent at that time,
showed how these could be incorporated into the full quantum theory.
The Higgs field configuration (which implemented the spontaneous breaking) also gave a complete nonsingular characterization of the relevant topological issues. Further, the mass of the monopole could be calculated as an unambiguous and finite quantity.
We should also note that a little later, Wu and Yang gave
a description of Dirac's monopole which highlighted the
connection to ideas of topology \cite{wu-yang}.

It was in this research milieu that Bal's work moved in a decisive way
to topological questions in field theory. He authored a number of papers on charged particle-monopole scattering, structure of gauge vacua, the Gribov problem, functional and Hamiltonian analyses of such questions, etc.
In this context, two papers deserve special mention.
The first of these, jointly with P. Salomonson, B-S Skagerstam and
J-O Winnberg, considered the description of charged particles interacting 
with nonabelian gauge fields \cite{BSSW}. The charge degrees of freedom (the
``color" degrees of freedom, so to speak) were
taken account of by Grassmann variables.
The equations of motion for the proposed Lagrangian were shown to
agree with the equations which had been proposed by Wong a few years earlier \cite{Wong}. Further, Bal {\it et al} set up the quantum theory and analyzed the
representations of color charges which would be obtained.
It should be mentioned that, almost simultaneously,
 a very similar theory was proposed and analyzed
 by A. Barducci, R. Casalbuoni and L. Lusanna \cite{barducci}.
Some earlier papers  by Brink {\it et al} 
constructed or used
supersymmetry  to discuss spinning particles \cite{brink}.
As Al Stern recalls,
Bal's sabbatical leave in Sweden in 1976 
also helped in the exchange of ideas with these research groups.
 
 The second paper, jointly with S. Borchardt and A. Stern,
 took these ideas further \cite{BBS}. The description of the color degrees of freedom
 was now given in terms of a topological action on a coset of the relevant Lie group. Once again, the canonical framework and quantization were discussed.
The topological action was in fact the co-adjoint orbit action
used for the construction of representations of Lie group in the Borel-Weil-Bott theory and in the work of Kirillov and Kostant and Souriau.
In its simplest version, the action takes the form
\beq
{S} = i\, {n\over 2} \int dt\, \Tr \left(\sigma_3 g^{-1} {d g\over dt }\right)
= i\, {n\over 2} \int dt\, \Tr \left(\sigma_3 g^{-1} {\dot g }\right)
\label{1}
\eeq
where $g$ is a $2\times 2$ unitary matrix
giving an element of $SU(2)$ and $\sigma_i$ are the Pauli matrices.
The action is defined on $SU(2)$, but the symplectic two-form 
corresponding to
Eq.(\ref{1}) is defined on $SU(2)/U(1)$.
This two-form is closed but not exact, 
$SU(2)$ can be viewed as a nontrivial fiber bundle with 
$U(1)$ (corresponding to
the $\sigma_3$-direction) as the fiber.
Quantum dynamics consistently descends to $SU(2)/U(1)$
provided $n$ is an integer; this is the Dirac quantization condition.
In retrospect, one may think of $S$ in Eq.({\ref{1}) as the ur-action from which
various Wess-Zumino-Witten (WZW)-type topological terms would arise.
For example, the expansion of the action in Eq. (\ref{1}) using 
star-products on a
noncommutative space naturally leads to the higher rank
differential forms which appear as WZW terms.
The two papers mentioned here made clear that this topological action was not
an esoteric mathematical curiosity but relevant to
everyday physics.
In a follow-up paper, Bal (and T.R. Govindarajan and B. Vijayalakshmi)
extended these ideas to constructing particles with spin 
within a classical theory \cite{BGV}.

This was the time that Bal's collaboration with his long term friend
Giuseppe (``Beppe") Marmo started flourishing.
(Bal had also been an informal guide to Beppe's laurea thesis.)
The co-adjoint orbit action is the quintessential and paradigmatic example of how to formulate quantum theories on a nontrivial fiber bundle.
With Beppe's expertise on fiber bundles, Bal and collaborators soon produced a number of papers on related but diverse topics. Charged particle-monopole interactions were analyzed without the use of Dirac strings \cite{BMS1}.
Dealing with the full fiber bundle $SU(2)$ rather than the base space
$SU(2)/U(1)$
would help evade the string singularities and this constituted the minimalist
formulation of the requisite topological features of the 't Hooft-Polyakov
monopoles.
This was soon followed by analyses of supersymmetric particles
\cite{BMSS1}
and of spinning particles in general relativity along similar lines
\cite{BMSS2}.

Bal's style of physics always involved a fair number of collaborators,
as he loved to discuss, often with his own attempts at 
explaining things helping to crystallize his thoughts 
and identify new avenues to be explored. 
A particularly fruitful case of a large collaboration involved
G. Marmo, N. Mukunda, J. Nilsson, E.C.G. Sudarshan and
sometimes F. Zaccaria and A. Simoni.
They produced some remarkable papers on the canonical structure of dynamics. One of these papers discussed the universal unfolding in 
Hamiltonian systems \cite{gang1}.
 The focus here was to look at the Hamiltonian structure of theories for which the symplectic two-form is a closed but not exact
two-form. The action in such cases would necessarily require an additional variable, a fiber direction, due to the nontrivial nature of the fiber bundle.
In the Hamiltonian analysis, the wave functions would involve
nonintegrable phase factors, very much along the lines of Dirac's original paper on monopoles. Although the terminology of nonintegrable phase factors
was used, essentially this meant that the wave functions are sections of a line bundle. 
In constructing an effective action to represent anomalies, Wess and Zumino had already noticed that an extra direction would be required \cite{WZ1}.
However, the fact that this leads to a symplectic form which is closed but not exact was not clear. In fact, the
deeper topological implications of Wess-Zumino work only became clear after
Witten's work on the global aspects of current algebra \cite{witten1}
and in his work
on nonabelian bosonization \cite{witten2}, in related work by Novikov \cite{novikov},
and later in the cohomological derivation of anomalies \cite{stora}.
Nevertheless, nontrivial fiber bundles and cohomology were in the air, and the
paper by Bal {\it et al} did anticipate and clarify certain facets of the problem.

Another remarkable result produced by the same gang was on color breaking
by monopoles. In any grand unified theory (GUT), a simple group is spontaneously broken to $H= \bigl[SU(3) \times SU(2) \times U(1)\bigr]/\C$,
$\C \sim \mathbb{Z}_6$, which gives the gauge group for the Standard Model.
By virtue of having nontrivial connectivity for the latter group,
there are necessarily monopoles.
In the simple case of the $SU(5)$ GUT,
the basic monopole can be viewed as an 't Hooft-Polyakov monopole in 
an $SU(2)$ subgroup broken to $U(1)$, where the $SU(2)$ straddles
the $SU(3)$ of color and $SU(2)$ of the electroweak theory.
As a result, any particular classical solution breaks the color symmetry.
However, the general expectation was that one could 
quantize the moduli of color rotations, viewed as collective coordinates
of the monopole, and that this would lead to monopoles in various
representations of the color group in the quantum theory.
Bal {\it et al} gave a topological argument, based on the realization
of asymptotic color transformations, that this is not the case, color
rotations cannot be unitarily realized \cite{gang2}.
Basically, to define the gauge potential on the sphere at
spatial infinity, we need at least two patches with a transition
function ($U(1)$ element) $\tau (\theta, \vf)$ on a collar around the equator.
A global color transformation $g$ does not leave $\tau$ invariant,
$g \, \tau (\theta, \vf) \, g^{-1} \neq \tau (\theta, \vf)$
because of the placement of the $SU(2)$ in $SU(5)$.
The same result was independently obtained at the same time 
by Manohar and Nelson \cite{nelson1}.
 As clarified soon after by Coleman and Nelson, the result could also be understood 
 as the divergence of the effective action for color
rotations due to the asymptotic behavior of monopole configurations
as required by topology \cite{nelson2}. (There were indications of this in the work of
Abouelsaood as well \cite{abouel}.)
This is a profound result applicable to any situation where the
unbroken subgroup $H$ is nonabelian and the monopole solutions
involve fields in the nonabelian part of $H$.
In some sense, it parallels the breaking of Lorentz invariance
in the charged sectors of quantum electrodynamics
due to the infrared photons \cite{strocchi}.
This is also a problem Bal returned to many times, 
showing that it is a general feature when the
wave functions are sections of a twisted bundle on the
configuration space and the group $H$ is nonabelian,
the ethylene molecule being another simple, and eminently physical,
 example \cite{ethylene1}.

\section{Anomalies and skyrmions}

My own work with Bal started around 1980. Although I participated in some early work on chiral models and on $\mathbb{Z}_2$-charged monopoles,
things moved into high gear with the work on anomalies or the quantum breaking of classical symmetries.
Already by the early 1970s, anomalies had been calculated for
Abelian and nonabelian currents using Feynman diagrams \cite{{Ad},{BJ},{Bar}}.
 Wess and Zumino
had discussed the representation of anomalies by an effective action
(as mentioned earlier) and had also given the consistency conditions such an action should obey \cite{WZ1}.
In 1979, Fujikawa gave a nice path-integral derivation of the axial
$U(1)$ anomaly, interpreting it as due to the Jacobian in the axial
$U(1)$ transformation of the measure for functional integration \cite{Fuji}.
The essence of the calculation was the functional trace,
$\Tr (\gamma_5)$.
Fujikawa's calculation was for Abelian currents although he did include
nonabelian background gauge fields.
Bal suggested that we could calculate the full nonabelian anomalies
{\it \`a la} Fujikawa, and my fellow student Garry Trahern
 and I were able to use heat kernel expansions to obtain the
 full result \cite{BMNT}, namely, the same expression as had been calculated by
 Bardeen \cite{Bar} by diagrammatic methods. However, there was another term
 which did not agree, but, interestingly, this could be shown to be removable by
 choice of regularization using the WZ consistency conditions, which
 Bal (and Beppe Marmo) had discussed. 
 
We went on to analyze the WZ effective action for anomalies \cite{BNT}.
The quantization of the coefficient
of this action was demonstrated for the Abelian anomalies.
As the simplest nonabelian extension, the case of $SU(2)$ was also analyzed,
 but the full topology and implications for physics eluded us.
 
In the late 1950s, T.H.R. Skyrme had proposed that baryons are topological solitons made of meson fields \cite{skyrme}.
This was a truly strange idea at that time.
Topology was not all that familiar to contemporary physicists, methods to include solitonic states in the quantum theory would not be developed
 until early 1970s, and no one knew how fermionic particles could be made from bosons.\footnote{The last bit may be a slight overstatement.
 A few years after Dirac's work,
 M.N. Saha had shown that the composite of a charged particle with the background of a monopole would acquire an extra half-unit of angular momentum.
 Here the monopole is described by a classical field, whereas
 Skyrme claimed the possibility of change of statistics
in the full quantum theory.
 Nevertheless, one could argue that already in Saha's paper
we can see intimations of how
 nontrivial topology could transmute spin or statistics \cite{saha}.} 
 Nevertheless, the idea intrigued many physicists
and the confirmation of the (1+1)-dimensional analog, namely, 
the sine Gordon-Thirring model equivalence \cite{sG-T}
(a variant of which had also been argued by Skyrme),
lent credence to his ideas.
After our work on anomalies, Bal felt that we could 
tackle skyrmions.
Bal's work with G. Trahern and A. Stern on formulating
nonlinear sigma models in gauge theory language
had also helped to prime his thinking along these lines
\cite{BTS2}.
And a recent paper by Pak and Tze  had certainly piqued our
interest \cite{PakTze}.

The Lagrangian for the
nonlinear model describing pseudoscalar mesons
is of the form
\beq
\L = -{f_\pi^2 \over 2} \Tr ( \del_\mu U \, U^\dagger 
 \del_\mu U \, U^\dagger )
 + {1\over 32 e^2} \Tr ( [\del_\mu U U^\dagger, \del_\nu U U^\dagger]^2)
 +\cdots
 \label{2}
\eeq
where we have also included the quartic term needed for the
classical stability of the solitons.\footnote{This term
is not needed quantum mechanically, the
uncertainty principle for the collective coordinates suffices to stabilize 
the soliton \cite{JSchS}.}
Here
$U$ is a field taking values in $SU(3)$ (for three flavors of quarks).
The expansion of $U$ around the identity would yield terms relevant to the dynamics of the pseudoscalar mesons, while
skyrmions are field configurations corresponding to nontrivial elements
of the homotopy group $\Pi_3 (SU(3)) = \mathbb{Z}$.
We introduced a coupling to baryons ($N$) of the form
\beq
\L_{\rm int} = -m \,\bigl[{\bar N}\, U (1-\gamma_5) N + h.c.\bigr]
\label{3}
\eeq
Bal, my fellow students Rajeev and Stern and I were then able to calculate
the baryon charge density in the background of $U$ as
\cite{BNRS}
\beq
\bra{U} N^\dagger N \ket{U} = {1\over 24 \pi^2}\epsilon^{ijk}
\Tr \left( \del_i U U^\dagger\,  \del_j U U^\dagger \,  \del_k U U^\dagger
\right) + {\rm total ~derivatives}
\label{4}
\eeq
The first term on the right side is the topological 
charge density, its integral giving the winding number
of $\Pi_3 (SU(3))$, i.e., the soliton number of the skyrmion.
This gave a clear and unambiguous proof that the soliton number
can be identified as the baryon number.
We also considered quantizing the moduli for the skyrmions
along the lines of the Goldstone-Jackiw approach to solitons
in field theory \cite{GolJ}, although the later approach of Adkins, Nappi and Witten 
was simpler (and more popular) \cite{ANW}.
As for the fermionic nature, we relied on the possibility that
skyrmions in the $SU(2)$ theory could be quantized
as fermions, as had been argued by Williams many years earlier,
again based on topological arguments and a wonderful theorem of Freudenthal's \cite{williams}.

Around this time, Witten came to give a seminar
in Syracuse. In his large $N_c$ analysis of QCD, he had argued that baryons
should be treated as solitons. This aligned with
Skyrme's idea and we had a number of discussions on this.
Clearly Witten was intrigued, and fairly quickly his remarkable
papers on
the global aspects of current algebra \cite{witten1} and on current algebra and baryons
\cite{witten3}
appeared. He was able to clarify the topological aspects of
the WZ effective action for anomalies and to prove that skyrmions 
would have half-odd-integer values for spin (for odd number of colors for quarks).
Witten also graciously acknowledged discussions with
us.

Bal continued to be interested in skyrmions
for several more years, producing both research 
articles and influential reviews.
A key development was about the dibaryon.
There is an embedding of $SU(2)$ 
in $SU(3)$, corresponding to the branching
$\underline{\bf 3}$ of $SU(3) \rightarrow
\underline{\bf 3}$ of $SU(2)$, which is different from the minimal
standard embedding, which corresponds to 
$\underline{\bf 3}$ of $SU(3) \rightarrow
\underline{\bf 2} \oplus \underline{\bf 1}$ of $SU(2)$.
Skyrmions of the nonminimal embedding
have baryon number equal to 2 and necessarily involve
strangeness, since the full $SU(3)$ is utilized in the configuration.
Bal (with A. Barducci, F. Lizzi, V.G.J. Rodgers and A. Stern)
pointed out the possibility of the nonminimal embedding
and  
calculated the mass ($\sim 2.2\,{\rm GeV}$) and analyzed
various other features of the dibaryon within this framework
\cite{dibaryon}.
This is an important contribution, since,
although the idea of dibaryons had been considered before in the 
bag model of hadrons (it would be a bound state of six quarks
$\sim uuddss$) \cite{jaffe}, the complementary
skyrmion picture is more amenable to calculations for certain features.
We may also note that dibaryons have been considered as a candidate for dark matter and evidence to date seems not to have excluded them
\cite{farrar}.

Today topology is ubiquitous in theoretical physics. Yet there are not many examples of topological effects with direct observational support.
Anomalies pass the test, their effect being directly measurable in
the $\pi^0 \rightarrow 2 \gamma$ decay and in the quantum hall effect.
The high mass of $\eta'$ relative to other pseudoscalar mesons
may even be taken as indirect evidence.
Skyrmions also pass the test, perhaps to a lesser
extent. After all  they are the well-known baryons, although an alternate description using quarks is possible.
It is interesting that the Syracuse group led by Bal
could make useful contributions to these
areas where the abstract mathematics of topology touches upon 
down-to-earth experimental evidence.

\section{The spin-statistics stream}

From the mid-1980s to the mid-1990s, a lot of Bal's efforts were devoted to
the spin-statistics connection and its many realizations.
The key point was to focus on the topology, specifically the first homotopy group $\Pi_1$, of the configuration space $\C$. While the 
spin-statistics connection has a long history, most of the early proofs
required relativistic invariance \cite{spinst1}.
There were however
indications that this was an overkill, and, in fact, Sudarshan had given
a simpler proof in 1968 without relativity \cite{sudar}, following the ideas of Schwinger \cite{schwinger2}.
Also Finkelstein and Rubinstein gave arguments using only
$\Pi_1 (\C)$ \cite{finkrub}, this was also used for solitons and generalized to geons 
by Friedmann and Sorkin \cite{sorkin1}. (For example, spin-statistics
for skyrmions would require such considerations.)

Bal and company (A. Daughton, Z-C Gu, G. Marmo, R. Sorkin and A.M. Srivastava) were able to give a general proof without relativity or field theory, but assuming the existence of antiparticles \cite{bal-daught}. The configuration space
consists of particle positions excluding coincidence and antiparticle
positions (again excluding coincidence) but with identifications
when a particle and antiparticle are at the same point reflecting
the annihilation of the pair.
To include spin, frames were also attached to the particles and
antiparticles. Then, by fairly simple arguments,
the exchange of particles could be shown to be reducible to
the no-exchange case with a $2\pi$ rotation of a spin-frame, leading to
the required spin-statistics relation. (The spinless case had been considered
along these lines by Tscheuschner \cite{Tscheu}.)
Even though this is at the level of particles and not field theory, this
result is very general,
since the ``particles" could even be extended objects,
quasiparticles or composites.
In fact, the result was soon extended to strings as well, with the homology of the
configuration space as the key quantity of interest \cite{bal-mcglinn}.

Bal was also fascinated by unusual cases of particle statistics.
It had been known for some time that point particles with associated
long range magnetic potentials, which could result from coupling
 to a Chern-Simons (C-S) term, could lead to fractional statistics in 2+1 dimensions \cite{wilcz}.
A nonabelian generalization was obtained (with M. Bourdeau and S. Jo)
with various types of exotic statistics for the sources \cite{bourjo}. This was also
extended to strings using a $BF$-theory, which is a string generalization 
of the C-S term \cite{stringcs}.
Bal and collaborators also obtained another fascinating result for curved spacetime: Geons in three dimensions do not have to obey any definite statistics, and one could have, for example, a situation where a tensorial geon obeys
Fermi-Dirac statistics \cite{stat-GR}.
Bal (with E. Ercolessi) even considered statistics on networks \cite{stat-net}.

This may also be an apt place to mention another work,
which I think is important but somewhat under-appreciated.
While we focused on statistics, the obverse of the same coin is
the discussion of spin in terms of the representations of the 
Poincar\'e group.
For  a particle, one can analyze these representations
using the appropriate version of the co-adjoint orbit action in Eq.(\ref{1}).
Bal (and F. Lizzi and G. Sparano) developed a similar description for
strings, enabling a uniform characterization and classification of
different types of strings \cite{BLS3}. In particular, null strings could be analyzed
in much the same way as massless particles.
(Extended objects had fascinated Bal for a long time, as evidenced by
his earlier papers on the subject (with G. Marmo, B-S. Skagerstam
and A. Stern) \cite{extended}.)

The whole body of work here is important for its generality and its role in highlighting the limitations and caveats of the spin-statistics 
theorem. Bal continued to be interested in these questions even as he moved on to noncommutative spaces. 

\section{Edge states}

In a gauge theory, the generator of gauge transformations
$G(\theta)$, smeared with test function $\theta$, will annihilate all physical states, and weakly commute with observables, for $\theta$ vanishing on
the boundary of the space under consideration.
$G(\theta)$, with test functions which do not vanish on the boundary
are physical observables and generate the edge states of the theory.
For $\theta$ constant at spatial infinity, these correspond to
global charge rotations, while for $\theta$ as a general function 
on the boundary, we get an infinity of physical states.
In the absence of any dynamics on the edge, the values
of $\theta$ are superselected.
During the 1960s and '70s, edge states were a seldom discussed topic in quantum field theory.
In the quantum Hall system, the dynamics of a droplet of electrons
is described by the (2+1)-dimensional Chern-Simons theory in the bulk,
on the edge are physical states corresponding to a chiral boson
\cite{QHE-review}.
More generally for bulk dynamics given by the nonabelian
Chern-Simons theory, the edge dynamics is described by the
WZW theory, as shown by Witten \cite{witten4}.
These results from the 1980s and ideas of holography generated interest in analyzing
edge states and the correspondence of bulk and edge dynamics,
and Bal and collaborators devoted considerable
effort to it.

Careful canonical quantization was shown to lead to the
currents obeying the Kac-Moody (KM) algebra on the edge
(work with G. Bimonte, K.S. Gupta, A. Stern) \cite{BBGupS}.
$BF$ theories, relevant to dimensions higher than 2+1, 
were also analyzed (with G. Bimonte and P. Teotonio-Sobrinho), 
identifying the symplectic form
for the edge states, effectively yielding
a generalization of the KM algebra \cite{teoton}.
A particularly interesting case, since it directly applies to the Hall system,
was the Maxwell-Chern-Simons (MCS) theory, analyzed with
T.R. Govindarajan, E. Ercolessi and R. Shankar \cite{BGES}.
If the space is taken to be a disk $D$,
the self-adjointness of the Hamiltonian allows for
general boundary conditions of the form
$^*dA = - \lambda \, A_\theta$ on the edge $\del D$, characterized by
the free real parameter $\lambda \geq 0$. Edge states exist only for
$\lambda = 0$.
Also, interestingly, already in 1995, long before entanglement in field theory
became fashionable, Bal (and his students L. Chandar and A. Momen)
analyzed the
entanglement entropy for the edge states of the MCS theory \cite{BChanM}.

The nature of boundary conditions required by self-adjointness of the
Hamiltonian and how it correlates to the existence of edge
states was also analyzed more generally (with L. Chandar and E. Ercolessi)
\cite{BChanEr}.
There were papers on edge states in general relativity, particularly relating to black hole horizons (with L. Chandar and A. Momen) \cite{BChanM2}.
The hope or motivation here was to see if there could
be a microscopic counting argument leading to
the Bekenstein-Hawking entropy in terms of such states,
an idea which has been pursued since by other researchers.
Another intriguing direction was in showing how variations in
boundary conditions could be viewed as topology change
\cite{BBM2}.
As an elementary example,
 two line segments could be viewed as two circles or a figure eight
depending on how the boundary conditions are implemented,
and boundary dynamics could go from one to the other.

Edge states continued to fascinate Bal for many years.
There were papers on topological insulators, 
chiral bags \cite{Asoreybal}, 
supersymmetric generalizations \cite{Asoreyach}, etc.
Bal and collaborators returned a few times to the
analysis of the superselection sectors of a gauge theory, defined by the boundary values of $\theta$, dubbing them as the ``sky group"
\cite{skygroup}.
Interestingly, different choices of $\theta$ on 
disconnected boundaries can lead to
a current in the bulk, a nonabelian version of the
Josephson current in a superconductor junction \cite{BNV1}.
Also an action for the edge dynamics with full electric-magnetic duality
was obtained \cite{edge-dual}.
While we may expect duality invariance for a theory like the
${\cal N} = 4$ SUSY YM, the point is that edge dynamics shows the duality invariance even without supersymmetry.
Edge states are also closely related to the issue of infrared divergences in
gauge theories, another problem of unwavering interest for Bal.
Among other things Bal and collaborators tried to come up with
testable consequences of the Lorentz violation for charged sectors mentioned earlier \cite{strocchi}, such as possible violations of the Landau-Yang theorem
\cite{asorey2}.
As Bal's interests expanded to include noncommutative geometry,
Bal (and K.S. Gupta and S. {K\"{u}rk\c{c}\"{u}o\v{g}lu}) also carried out the analysis of
edge states in the noncommutative version of the Chern-Simons theory
\cite{edge-NCCS}.

\section{Noncommutative and fuzzy geometry}

Bal devoted almost half of his scientific career, from the mid-1990s onwards, to the topic of noncommutative and fuzzy spaces.
To contextualize his contributions, let us note that a quantum theory of gravity has long been the most daunting challenge in theoretical physics.
Attempts to address this problem have led to
many lines of exploration and many avenues of thought over the last few decades.
String theory, including its avatar in terms of holography, could be argued to be
the most viable \cite{string}, but there are other possibilities as well. These include
loop quantum gravity \cite{loopg}, noncommutative geometry
\cite{ncgeom}, causal set theory
\cite{causal}, or even the
possibility that gravity is no true force at all, but a manifestation of
entropic considerations \cite{entropicg}.

The simplest case of a noncommutative manifold is the Moyal space
with
\beq
[x^\mu, x^\nu ] = i \, \theta^{\mu\nu},
\label{5}
\eeq
for a constant antisymmetric matrix $\theta^{\mu\nu}$,
where $x^\mu$ denotes the coordinates on the manifold.
The algebra of functions on such a space is characterized by the
(noncommutative) star product
\beq
( f* g) (x) = \exp\left( {i \over 2} \theta^{\mu\nu}{\del \over \del y^\mu}
{\del \over \del z^\nu}\right) \, f(y) \, g(z) \bigg\vert_{y=z=x}
\label{6}
\eeq
Similarly, in fuzzy geometry, functions on a manifold 
(which encode the geometry) are replaced by operators
on an $N$-dimensional Hilbert space $\H_N$, with the resultant
algebra of functions displaying noncommutativity $\sim {1\over N}$.
A Dirac operator or a Laplace operator is needed to carry information
about the emergent metric at large $N$.
The main attraction for such structures is that, the
noncommutativity, say via the associated uncertainty principle,
can smear out short distance singularities, thus potentially
avoiding problems of nonrenormalizability for gravity.
Fuzzy models are also amenable to numerical calculations.

The work of Bal and collaborators on this 
topic falls into four or five broad
lines of development. 
Of course, it should be kept in mind that
in such matters of general interest, there are always
many research groups working on similar things and
so a certain amount of mutual interaction
is involved in the advancement of the field.
Only the injection of key ideas into an evolving stream
of developments or the impactful analyses by Bal
and collaborators is all that we can try to highlight.

With a constant $\theta^{\mu\nu}$, the standard realization of Poincar\'e
symmetry is not obtained, since $x$'s on the left hand side
of Eq.(\ref{5}) will transform while the right hand side remains constant.
But the work by several authors had shown that there is a deformation
of the co-product (i.e, with a ``Drinfel'd twist") needed for the action of 
Lorentz transformation of products of operators which
allows for the realization of the full Poincar\'e symmetry
\cite{twist-P}.
Bal and collaborators (A.Pinzul, B.A Qureshi, and also
with T.R. Govindarajan, G. Mangano, A. Pinzul,  B.A. Qureshi and S. Vaidya)
showed that this leads to two important consequences.
The Drinfel'd twist, when combined with the
covariance of the identity of particles (i.e., identical particles remain identical even after a Lorentz transformation), is equivalent to redefining the
annihilation operator $c_p$ for a scalar field
as
\beq
c_p \, \Longrightarrow\, a_p = c_p \, e^{ {\half} p_\mu \theta^{\mu\nu} {\hat P}_\nu}
\label{7}
\eeq
where $p$ is the momentum label for the operator
and ${\hat P}$ is the momentum operator.
As a result, particle statistics is affected and this allows for the possibility of
Pauli-forbidden transitions in field theories on the Moyal space \cite{BGMPQV}.
The data on such Pauli-forbidden transitions would also allow for extracting bounds on $\theta$ from experiments;
in later work, Bal (and P. Padmanabhan and A. Joseph) argued that
$\theta \sim (10^{24} {\rm TeV})^{-2}$ \cite{BPJ}.
The possibility of obtaining unitary time-evolution in case of space-time noncommutativity (i.e., when $\theta^{0i} \neq 0$) was also explored
\cite{BGMT-S}.

Another related peculiarity is the following.
For the given commutation rule Eq.(\ref{5}) for the coordinates, 
the star product
in Eq.(\ref{6}) is just one possibility for the twist 
 in the algebra of the functions;
 it is possible to have different definitions keeping
 Eq.(\ref{5}) unaltered.
 While the corresponding algebras remain isomorphic, 
 and the classical theories also remain equivalent,
in the quantum theory they can have very different 
physical consequences.
This feature, which is an important point to pay attention to
for setting up field theories on noncommutative
spacetimes, was noted by Bal
and collaborators (A.Ibort, G. Marmo and M. Martone)
and analyzed in detail in a number of papers \cite{B-Ib-MM}.

Secondly, if we consider a $g^2\phi^4/4!$ theory, with the
noncommutative version $g^2\phi*\phi*\phi*\phi/4!$, a simple one-loop calculation
shows that the two-point vertex function gets a contribution
\beqar
\Gamma^{(2)} &=& {g^2 \over 96\pi^2}\left( \Lambda_{\rm eff}^2
- m^2 \log (\Lambda_{\rm eff}^2/m^2) + \cdots \right)\nonumber\\
\Lambda_{\rm eff}^2 &=&{1\over {(1/\Lambda^2) - p_\mu p_\nu (\theta^2)^{\mu\nu}}}
\label{8}
\eeqar
where $\Lambda$ is a UV cutoff.
If $\theta$ is set to zero (or if we take $p \rightarrow 0$ first), we get the expected quadratic 
UV-divergence as $\Lambda \rightarrow \infty$,
since $\Lambda_{\rm eff} = \Lambda$.
However, if we first take $\Lambda \rightarrow \infty$, the result is finite
but acquires an IR-divergence as $p \rightarrow 0$
\cite{UV-IR}. This is famous (or infamous) UV-IR mixing.
With the twisted realization of Poincar\'e symmetry, Bal and company showed that this UV-IR mixing can be avoided \cite{BPQur}.
(Oeckl had also obtained this result, see \cite{twist-P}.)
The work on field theories on Moyal space continued in several more
papers, analyzing discrete symmetries and other features.

Another important result was to show the absence of fermion doubling.
While the Nielsen-Ninomiya theorem shows that
fermion doubling is unavoidable in the usual discretization of a theory \cite{NielN},
this is not the case on fuzzy spaces.
On a fuzzy two-sphere, for example, since functions correspond to
matrices, one can define $SU(2)$ operators $L^{(l)}_i$,
$L^{(r)}_i$ with left and right actions on such ``functions". An associated  discrete
Dirac operator $\D = \sigma\cdot (L^{(l)}- L^{(r)}) + 
\mathbb{1}$ was used by Grosse, Klim{\v c}{\'i}k, and Pre{\v s}najder
in their work on $S^2_F$ \cite{GKP}.
 While this does not admit a chirality
operator,
Bal and collaborators (T.R. Govindarajan and B. Ydri) showed 
that one could project out the highest level eigenstates 
of $\D$ to be able to define
a chirality operator
and further, that this could be done in a way 
maintaining the continuum limit
and that it would also avoid fermion doubling \cite{BGY1}.
In a later paper (with Immirzi), a different Dirac operator
 (with the same continuum limit) was used \cite{BImm}.
Both cases, it was shown, led to
the Ginsparg-Wilson algebra, which is the crucial
structure for eliminating doubling.

Since nonlocality can be used to evade the Nielsen-Ninomiya 
theorem, and since fuzzy spaces necessarily 
have some nonlocality, this result may not seem surprising.
But the key point here is that one achieves a truncation of the theory to
a finite number of modes preserving various isometries and without
fermion doubling. Thus it has the potential to
give a better strategy for numerical simulations.

In any systematic approach to fields on a fuzzy space, which implies discretization and finite number of modes,
we encounter the question of how to deal with 
features associated to topology in continuum physics.
These include monopoles, solitons, chiral anomaly, etc.
Differential forms can be defined on fuzzy spaces, the analog of differentials being given by commutators with a suitable
projected version of the unimodular Dirac operator $\D /\vert \D \vert$.
The same operator via commutators or anticommutators plays the role
of the exterior derivative. Gauge fields, connections and curvatures can then be defined. Similar to the spectral action of Connes, the Dixmier trace
of products of curvatures or derivatives of fields can be used to define actions and topological quantum numbers \cite{ncgeom}.
 In this way, Bal (and S. Baez, S. Vaidya and B. Ydri)
analyzed topological features for fields on fuzzy spaces \cite{BBVY}.
The analysis was carried out using cyclic cohomology, which is the appropriate structure for these questions.
(There is also independent work by other groups on these issues,
see \cite{GKP} and \cite{Klim}.)

In another broad stream of work, Bal and collaborators worked on setting up
the nuts and bolts of fuzzy spaces. These included defining the fuzzy versions
of complex projective spaces, $\mathbb{CP}^2_F$ (with G. Alexanian, G. Immirzi, B. Ydri) \cite{CP2F}, more generally, $\mathbb{CP}^N_F$ (with
B. Dolan, J-H Lee, X. Martin and D. O'Connor) \cite{CPNF}, including supersymmetric versions (with S. {K\"{u}rk\c{c}\"{u}o\v{g}lu}, E. Rojas) \cite{fuzzysusy}, and working out the relevant
star products and so on. 
(There were different but related results by other groups as well \cite{others}.)
There was also work on
extracting the continuum limit of actions on fuzzy spaces (with
X. Martin and D. O'Connor) \cite{contlim}.
Another interesting paper interpreted the construction of
fuzzy spaces in terms
of Hopf algebras and explored some of the consequences \cite{BSec}.
There is certainly more to be investigated here.

Although a little out of historical
sequence, this is thematically a good place to point out that
Bal's first attempts towards noncommutative geometry 
predate most of the work discussed in this section.
In the mid-1990s, he and collaborators
(G. Bimonte, E. Ercolessi, G. Landi, F. Lizzi, G. Sparano and P.
Teotonio-Sobrinho) analyzed partially order sets or posets.
These are approximations to a manifold by a finite number of points,
with an ordering relation, an idea due to
Sorkin \cite{poset-sor}. (They are also related to the more general
causal set theory approach mentioned earlier \cite{causal}.)
The framework allows one to capture 
topological features of the manifold unlike a
simple-minded lattice approximation.
The algebra of functions on such a set is generally a
noncommutative $C^*$-algebra, so we could phrase this as
a sort of noncommutative lattice approximation.
Bal and collaborators analyzed physics on posets, 
taking it to a point where one could contemplate
numerical simulations \cite{poset-bal}.

\section{Anomalies, entanglement again}

It is presumably clear by now that quantum anomalies have been a
{\it leit motif} of Bal's career.
In yet another stab at this, Bal (and A. de Queiroz)
argued that an anomalous symmetry $G$ can be implemented
in the quantum theory by restricting the algebra of observables
$\A$ to a set invariant under $G$, at the cost of using mixed states
\cite{BQueir1}.
The full impact of this work may not be clear yet, perhaps it may
help eliminate that {\it b\^ete noire} of QCD, namely, strong
$CP$-violation via the $\theta$-angle \cite{BGQueir}.

We have already mentioned the global color problem for monopoles 
\cite{gang2} and
Bal's interest in entanglement entropy in the context of the MCS theory
\cite{BChanM}.
It was inevitable these would come together in the
ambience of increased interest in entanglement in the research community.
There are situations where there are different
density matrices (or states
in the GNS sense) which lead to the same observables,
i.e., $\Tr (\rho_1 a) = \Tr (\rho_2 a)$ for all $a \in \A$.
In such a situation, there is ambiguity in how the entropy of the
quantum state is defined.
Bal (and A. de Queiroz and S. Vaidya) analyzed these possibilities
within the framework of GNS construction of representations
of $\A$, with the global color issue and the ethylene molecule
as specific examples \cite{BQueirV}.

Entanglement is generally characterized in terms of a reduced density matrix,
but if, for argument's sake,
 we were to be puritanical requiring the algebra of observables
as the  starting point, how do we formulate entanglement?
Bal (and T.R. Govindarajan, A. de Queiroz, A.F. Reyes-Lega)
formulated and analyzed this question, showing how the reduced state
is obtained upon restriction of $\A$ to a smaller set
\cite{BGQueirReyes}.
Again, these are matters whose import in quantum physics is 
not fully understood yet.

\section{Seeking verifiability}

Bal's early work in physics started with issues close to experiment,
such as pion-nucleon scattering, and remained so, to some extent,
through the early years of skyrmion work.
As high energy physics grew distant from easy and direct verifiability,
as with the field, Bal's work moved into more mathematical 
areas. 
But the desire to connect with matters closer to experiment was
always there, he emphasized its importance
and always kept an eye out for any such
possibility. 
I would like to highlight a few forays
of this nature.

Bal, with B. Rai, G. Sparano and A.M. Srivastava, considered
modifying the chiral Lagrangian of QCD to take account of
the scale anomaly \cite{isospinballs}.
They showed that the theory could then admit
nontopological solitons, which are stable
if the isospin exceeds a certain critical value. They
also discussed the possibility of observing such ``isospin balls"
in heavy ion collisions.

We have already mentioned Pauli forbidden transitions
for physics in Moyal spaces and how they help to put an experimental
bound on $\theta^{\mu\nu}$, which characterize the noncommutativity
\cite{BPJ}.
Earlier, Bal and E. Akofor, S. Jo, A . Joseph and B. Qureshi
had argued that a nonzero $\theta^{\mu\nu}$ could lead to anisotropies
in the cosmic microwave background radiation and put limits
on $\theta^{\mu\nu}$ from cosmological data \cite{ABJJQu}.

Returning to the old problem of QCD, Bal, de Queiroz and 
Vaidya argued that a consistent truncation of (3+1)-dimensional
Yang-Mills theory, which still carries many of
the topological features of the full theory, is given by a matrix
model with the Hamiltonian
\beq
H = -{g^2 \over 2 R} {\del^2\over \del M_{ia}^2} + {1\over 2 g^2}
 \left[ \Tr (M^TM) - 6\, \det M + (\Tr (M^TM))^2 - \Tr (M^T M M^T M)
 \right]
 \label{9}
 \eeq
 where $M_{ia}$ is a $3\times 3$ matrix, with spatial
 rotations ($R$) and $SO(3)$ gauge transformations
 $g_{ab}$
 acting as $M_{ia} \rightarrow R_{ij} M_{jb} g_{ab}$ \cite{BQueirV2}.
 (This is for the gauge group $SU(2)$.)
This gives another finite mode approximation
similar to lattice gauge theory; however, it is
not only tractable by numerical simulations but also
preserves many topological features. Variational analyses and
numerical simulations carried out
with N. Acharyya, M. Pandey, S. Sanyal and S. Vaidya
seem to give reasonable glueball and hadron spectra \cite{Ach}.

Another important observation is again about QCD.
It has been known for quite some time that QCD can have a color-flavor
locked (CFL) phase at high baryon
number density, due to a pairing effect similar to what happens in a superconductor. The expectation value of the relevant order parameter
is unchanged under a common color and flavor
transformation, hence the CFL nature.
The symmetry group $G$ acting on the field (corresponding to the order
parameter) and the unbroken subgroup $H$ are given by
\beq
G = {SU(3)_{\rm C} \times SU(3)_{\rm F} \times U(1)_{\rm B}
\over \mathbb{Z}_3 \times \mathbb{Z}_3}, \hskip .2in
H= {SU(3) \times \mathbb{Z}_3
\over \mathbb{Z}_3 \times \mathbb{Z}_3}
\label{10}
\eeq
Bal and S. Digal and T. Matsuura observed that since 
$G/H = U(3)$ and $\Pi_1 (U(3)) = \mathbb{Z}$,
this phase allows for strings, the elemental ones carrying
nonabelian fields or fluxes \cite{BDMat}.
They also worked out many properties of such strings.
This is an important observation which goes towards
elucidating the nature and properties of the phase diagram of QCD.

\section{The less tangible}

In the theoretical sciences, while research articles are evidently the concrete expression of an individual's contributions, they do
not capture the totality of his/her impact.
Although less quantifiable, students and collaborators are 
important in the larger domain of influence. 
A glance at the articles and the list of contributors to the Bal Festschrift
in 2023 will suffice to convey a sense of this significant aspect of Bal's life
\cite{festschrift}.
As Bal himself noted recently:
``{\it I have contributed a few significant results to quantum theory, but my enduring contributions have been to the training of young researchers}".
My fellow students and I continue, to varying degrees, to exercise
the Bal-way-of-thinking about physics, in how to formulate and tackle a problem, as well as in making value judgements of what is more and what is less important.
The ``316" meetings were an integral part of Bal's and our life.
Every afternoon in Syracuse, Bal, his students, collaborators and visiting scientists would meet in Room 316 of the Physics Building, discussing
any and all problems in physics. 
I think all participants, including casual drop-ins, shared the 
sense of value and appreciation of this, as articulated in the lovely article by Fedele Lizzi
 \cite{fedele}.
Even after his retirement Bal continued this tradition via online meetings,
until a few weeks before his passing.

Along the same lines, Bal's many books brought in a larger audience to the fold. Particularly noteworthy among his seven books are
{\it Gauge Symmetries and Fibre Bundles- Applications to Particle Dynamics} and
 {\it Classical Topology and Quantum States}, both jointly
with G. Marmo, B.S. Skagerstam and A. Stern \cite{{bk-BMSS1},{bk-BMSS2}}, which
systematized the emerging role of topology in physics.
The {\it Lectures on Fuzzy and Fuzzy Susy Physics}, with
S. {K\"{u}rk\c{c}\"{u}o\v{g}lu} and S. Vaidya \cite{bk-BKV}, played a similar role
for fuzzy physics.
Also his {\it Group Theory and Hopf Algebras}
(with S. Jo and G. Marmo) \cite{BalJM}, and its antecedent {\it Lectures on Group Theory for Physicists},
with C.G. Trahern, have been go-to references for a number of courses on
these topics \cite{BalT-group}.

Bal also served as the managing editor of the International Journal of Modern
Physics A and Modern Physics Letters A as well as on the advisory boards of several institutions.
In recognition of his scientific work, Bal received the Chancellor's
Citation for Excellence and the Wasserstrom Prize for Excellence
in Teaching from Syracuse University. He was also awarded a prize
by the US Chapter of the Indian Physics Association and, of course, was
elected a Fellow of the American Physical Society.

In the end, let us note that a man is a complex irregular polyhedron and all we can hope to do is
illuminate one or two facets as we all bring our own slants to how we hold
the light. Yet, even with the limited view I have managed
to present, we can say with certainty that Bal has
had an influential and richly productive life.
\section*{Acknowledgments}
I gratefully acknowledge useful input from
Bal's students, collaborators and associates
in preparing this article, especially Manuel Asorey, 
Thupil Govindarajan, Kumar Gupta, Dimitra Karabali, Seckin
{K\"{u}rk\c{c}\"{u}o\v{g}lu}, Fedele Lizzi, Denjoe O'Connor, Aleksandr Pinzul,
Vincent Rodgers, Bo-Sture Skagerstam, Rafael Sorkin, Allen Stern and
S. Vaidya.

This work was supported in part by the U.S. National Science Foundation Grant No. PHY-2412479.




\begin{thebibliography}{99}
\bibitem{weinberg} S. Weinberg, \PRL~{\bf 17}, 616 (1966).\\
(This work also synthesized contributions from many researchers,
including Y. Nambu, M. Gell-Mann, F. G\"ursey, M. Goldberger and
S. Treiman, S. Adler, W. Weisberger and others.)

\bibitem{gundzik} A.P. Balachandran, M.G. Gundzik and F. Nicodemi, 
Il Nuovo Cimento {\bf 44A}, 1257 (1966); \NP~{\bf B 6}, 557 (1968).

\bibitem{dirac} P.A.M. Dirac, Proc. Roy. Soc. {\bf A 133}, 60 (1931).

\bibitem{dirac2} P.A.M. Dirac, \PR~{\bf 74}, 817 (1948).

\bibitem{schwinger1} J. Schwinger, \PR~{\bf 144}, 1087 (1966);
\PR~{\bf 151}, 1048 (1966); \PR~{\bf 151}, 1055 (1966); \PR~{\bf 173}, 1536 (1968); Science {\bf 165}, 757 (1969); Science {\bf 166}, 690 (1969).

\bibitem{zwan} D. Zwanziger, \PR~{\bf 176}, 1480 (1968);
\PR~{\bf 176}, 1489 (1968); \PR~{\bf D 3}, 880 (1971).

\bibitem{HP} G. ’t Hooft, Nucl. Phys. {\bf B 79}, 276 (1974); A.M. Polyakov, JETP Lett. {\bf 20}, 194 (1974).

\bibitem{wu-yang} T.T. Wu and C.N. Yang, \PR~{\bf D 12}, 3845 (1975);
\NP~{\bf B 107}, 365 (1976).

\bibitem{BSSW} A.P. Balachandran, P. Salomonson, B-S. Skagerstam and J-O. Winnberg, \PR~{\bf D 15}, 2308 (1977).

\bibitem{Wong} S. K. Wong, {Nuovo Cim.} {\bf 65 A}, 689 (1970)

\bibitem{barducci} A. Barducci, R. Casalbuoni and L. Lusanna,
{Nuovo Cim.} {\bf A 35}, 377 (1976); \NP~{\bf B 124}, 93 (1977).

\bibitem{brink} L. Brink {\it et al}, \PL~{B 64}, 435 (1976);
L. Brink, P. di Vecchia and P. Howe, \NP~{\bf B 118}, 76 (1977).

\bibitem{BBS} A.P. Balachandran, S. Borchardt and A. Stern, \PR~{\bf D 17}, 3247  (1978).

\bibitem{BGV} A.P. Balachandran, T.R. Govindarajan and B. Vijayalakshmi, \PR~{\bf D 18}, 1950 (1978).

\bibitem{BMS1} A.P. Balachandran, G. Marmo, B-S. Skagerstam and A. Stern, \NP~{\bf B 162}, 385 (1980).

\bibitem{BMSS1} A.P. Balachandran, G. Marmo, B-S. Skagerstam and A. Stern, \NP~{\bf B 164}, 427 (1980).

\bibitem{BMSS2} A.P. Balachandran, G. Marmo, B-S. Skagerstam and A. Stern, \PL~{\bf B 89}, 199 (1980).

\bibitem{gang1} F. Zaccaria, E.C.G. Sudarshan, J.S. Nilsson, N. Mukunda,
G. Marmo and A.P. Balachandran, \PR~{\bf D 27}, 2327 (1983).

\bibitem{WZ1} J. Wess and B. Zumino, \PL~{\bf B 37}, 95 (1971).

\bibitem{witten1} E. Witten, \NP~{\bf B 223}, 422 (1983).

\bibitem{witten2} E. Witten, \CMP~{\bf 92}, 455 (1984).

\bibitem{novikov} S.P. Novikov, Usp. Mat. Nauk. {\bf 37},
3 (1982).

\bibitem{stora}  B. Zumino, Les Houches Lectures, 1983, reprinted in
S.B. Treiman et al, {\it Current Algebra and Anomalies}, Princeton University
Press (1986); R. Stora, Lectures at the Carg\`ese Summer Institute on
Progress in Gauge Field Theory, 1983; B. Zumino, Y-S. Wu and A. Zee,
Nucl. Phys. {\bf B 239}, 477 (1984).

\bibitem{gang2} A.P. Balachandran, G. Marmo, N. Mukunda, J.S. Nilsson,
E.C.G. Sudarshan and F. Zaccaria, \PR~{\bf D 29}, 2919, (1984);
\PR~{\bf D 29}, 2936 (1984).

\bibitem{nelson1} P.C. Nelson and A. Manohar, \PRL~{\bf 50}, 943 (1983)

\bibitem{nelson2} P.C. Nelson and S.R. Coleman, \NP~{\bf B 237}, 1 (1984).

\bibitem{abouel} A. Abouelsaood, \NP~{\bf B 226}, 309 (1983);
\PL~{\bf B 125}, 467 (1983).

\bibitem{strocchi} J. Fr\"ohlich, G. Morchio and F. Strocchi, 
Ann. Phys. {\bf 119}(2), 241 (1979);
\PL~{\bf B89}, 61 (1979); J. Fr\"ohlich, \CMP~{\bf 66}, 223 (1979);
D. Buchholz, \CMP~{\bf 85}, 49 (1982);
\PL~{\bf B174}, 331 (1986).

\bibitem{ethylene1} A. P. Balachandran, A. Simoni, and D. M. Witt, Int. J. Mod. Phys. {\bf A7}, 2087 (1992).

\bibitem{Ad} S. Adler, \PR~{\bf 177}, 2426 (1969); 
S. Adler and W.A. Bardeen, \PR~{\bf 182}, 1517 (1969).

\bibitem{BJ} J. Bell and R. Jackiw, Nuov. Cim. {\bf 60A}, 47 (1969).

\bibitem{Bar} W.A. Bardeen, \PR~ {\bf 184}, 1848 (1969).

\bibitem{Fuji} K. Fujikawa, \PRL~{\bf 42}, 1195 1979.

\bibitem{BMNT} A.P. Balachandran, G. Marmo, V.P. Nair and C.G. Trahern, \PR~{\bf D 25}, 2713 (1982).

\bibitem{BNT} A.P. Balachandran, V.P. Nair and C.G. Trahern, 
\PR~{\bf D 27}, 1369 (1983).

\bibitem{skyrme} T.H.R. Skyrme, Proc.
Roy. Soc. {\bf A 260}, 127 (1961); Nucl. Phys. {\bf 31}, 556 (1962); J. Math. Phys. {\bf 12}, 1735 (1970).

\bibitem{saha} M.N. Saha, Indian J. Phys. {\bf 10}, 145 (1936).

\bibitem{sG-T} S.R. Coleman, \PR~{\bf D 11}, 2088 (1975).

\bibitem{BTS2} A.P. Balachandran, A. Stern and C.G. Trahern,
\PR~{\bf D 19}, 2416 (1979).

\bibitem{PakTze} N.K. Pak and H.Ch. Tze, Ann. Phys. {\bf 117}, 164 (1979).

\bibitem{JSchS} P. Jain, J. Schechter and R.D. Sorkin,
\PR~{\bf D 39}, 998 (1989);
\PR~{\bf D 41}, 3855 (1990).

\bibitem{BNRS} A.P. Balachandran, V.P.
Nair, S.G. Rajeev and A. Stern, \PRL~{\bf 49}, 1124 (1982); \PR~{\bf D 27}, 1153 (1983).

\bibitem{GolJ} J. Goldstone and R. Jackiw, \PR~{\bf D 11}, 1486 (1975);
R. Jackiw, \RMP~{\bf D 49}, 681 (1977).

\bibitem{ANW} G.S. Adkins, C.R. Nappi and E. Witten,
\NP~{\bf B 228}, 552 (1983).

\bibitem{williams} J.G. Williams, J. Math. Phys. {\bf 11}, 2611 (1970).

\bibitem{witten3} E. Witten, \NP~{\bf B 223}, 433 (1983).

\bibitem{dibaryon} A.P. Balachandran, A. Barducci, F. Lizzi, V.G.J. Rodgers and A. Stern, \PRL~{\bf 52}, 887 (1984);
A.P. Balachandran, F. Lizzi, V.G.J. Rodgers and A. Stern, 
\NP~{\bf B 256}, 525 (1985).

\bibitem{jaffe} R. L. Jaffe, \PRL~{\bf 38}, 195 (1977); V. A.
Matveev and P. Sorba, Nuovo Cim. {\bf 45A}, 257 (1978).

\bibitem{farrar} G.R. Farrar, Z. Wang and X. Xu, arXiv:2007.10378[hep-ph]; G.R. Farrar and Z. Wang, arXiv:2306.03123[hep-ph].

\bibitem{spinst1} W. Pauli, \PR~{\bf 58}, 716 (1940);
G. Luders and B. Zumino, \PR~{\bf 110}, 1450 (1958);
N. Burgoyne, Nuovo Cim. {\bf VIII}(4), 607 (1958).

\bibitem{sudar} E.C.G. Sudarshan, in {\it Proceedings of the Nobel Symposium
8}, N. Svartholm (ed.), Almquist and Wiksell, Stockholm 1968,
pp. 379-386; Stat. Phys. Suppl. J. Indian Inst. Sci., June 123-137 (1975);
for a review with many earlier references, see
I. Duck and E.C.G. Sudarshan, Am. J. Phys. {\bf 66} (4), 284 (1998).

\bibitem{schwinger2} J. Schwinger, \PR~{\bf 82}, 914 (1951).

\bibitem{finkrub} D. Finkelstein and J. Rubinstein, J. Math. Phys.
{\bf 9}, 1762 (1968).

\bibitem{sorkin1} J.L. Friedman and R.D. Sorkin, \CMP~{\bf 89},
483 (1983); \CMP~{\bf 89}, 501 (1983);
R.D. Sorkin, \CMP~{\bf 115}, 421 (1988).

\bibitem{bal-daught} A.P. Balachandran, A. Daughton, Z.C. Gu, G. Marmo,
R.D. Sorkin and A.M. Srivastava, Mod. Phys. Lett. {\bf A 5}, 1575 (1990);
Int. J. Mod. Phys. {\bf A 8}, 2993 (1993);
Phys. Scripta T {\bf 36}, 253 (1991) (This was an invited talk by Bal
at the Nobel Symposium 79: The Birth and Early Evolution of
Our Universe, Sweden, June 1990, proceedings edited by J.S. Nilsson, B. Gustafsson, and B-S. Skagerstam.)

\bibitem{Tscheu} R.D. Tscheuschner, Int. J. Th. Phys. {\bf 28}, 1269
(1989).

\bibitem{bal-mcglinn} A.P. Balachandran, W.D. McGlinn, L. O'Raifeartaigh,
S. Sen and R.D. Sorkin, Mod. Phys. Lett. {\bf A 7}, 1427 (1992);
Int. J. Mod. Phys. {\bf A 7}, 6887 (1992); Int. J. Mod. Phys.
{\bf A 9}, 1395 (1994) [Erratum].

\bibitem{wilcz} F. Wilczek, \PRL~{\bf 48}, 1144 (1982);
\PRL~{\bf 49}, 957 (1982); R. Jackiw and A.N. Redlich, 
\PRL~{\bf 50}, 555 (1983).

\bibitem{bourjo} A.P. Balachandran, M. Bourdeau and S. Jo,
Mod. Phys. Lett. {\bf A 4}, 1923 (1989);
Int. J. Mod. Phys. {\bf A 5}, 2423 (1990);
Int. J. Mod. Phys. {\bf A 5}, 3461 (1990) [Erratum].

\bibitem{stringcs} C. Aneziris, A.P. Balachandran, L. Kauffman and A.M. Srivastava, Int. J. Mod. Phys. {\bf A 6}, 2519 (1991).

\bibitem{stat-GR} C. Aneziris, A.P. Balachandran, M. Bourdeau, S. Jo,
T.R. Ramadas and R.D. Sorkin, Mod. Phys. Lett. {\bf A 4}, 331 (1989);
Int. J. Mod. Phys. {\bf A 4}, 5459 (1989).

\bibitem{stat-net} A. P. Balachandran and E. Ercolessi,
Int. J. Mod. Phys. {\bf A 7}, 4633 (1992).

\bibitem{BLS3} A.P. Balachandran, F. Lizzi and G. Sparano,
\NP~{\bf B 277}, 359 (1986).

\bibitem{extended} A.P. Balachandran, B-S. Skagerstam and 
A. Stern, \PR~{\bf D 20}, 439 (1979);
A.P. Balachandran, G. Marmo, B-S. Skagerstam and 
A. Stern, J. Phys. G {\bf 7}, 1001 (1981).

\bibitem{QHE-review} See, for example, R.E. Prange and S.M. Girvin, {\it The Quantum Hall Effect},
2nd ed. (Springer-Verlag, Berlin, 2012); Z.F. Ezawa,
{\it Quantum Hall Effects} (World Scientific, Singapore, 2008);
T.H. Hansson, M. Hermanns, S.H. Simon and S.F. Viefers, \RMP~{\bf 89}, 025005 (2017);
D. Tong, {\it Lectures on quantum Hall effect}, arXiv:1606.06687[hep-th].

\bibitem{witten4} E. Witten, \CMP~{\bf 121}, 351 (1989).
See also M. Bos and V.P. Nair, \PL~{\bf B 223}, 61 (1989);
S. Elitzur, G. Moore, A. Schwimmer and N. Seiberg,
  \NP~{\bf B 326}, 108 (1989);
J. Labastida and A. Ramallo, \PL~{\bf B 228}, 214 (1989);
  A. Polychronakos, Ann. Phys. {\bf 203}, 231 (1990).

\bibitem{BBGupS} A.P. Balachandran, G. Bimonte, K.S. Gupta and
A. Stern, Int. J. Mod. Phys. {\bf A 7}, 4655 (1992);
{\bf A 7}, 5855 (1992).

\bibitem{teoton} A.P. Balachandran and P. Teotonio-Sobrinho,
Int. J. Mod. Phys. {\bf A 8}, 723 (1993);
A.P. Balachandran, G. Bimonte and P. Teotonio-Sobrinho,
Mod. Phys. Lett. {\bf A 8}, 1305 (1993).

\bibitem{BGES} A.P. Balachandran, T.R. Govindarajan, E. Ercolessi and
R. Shankar, Int. J. Mod. Phys. {\bf A 9}, 3417 (1994).

\bibitem{BChanM} A.P. Balachandran, L. Chandar and A. Momen,
 Int. J. Mod. Phys. {\bf A 12}, 625 (1997).

\bibitem{BChanEr} A.P. Balachandran, L. Chandar and E. Ercolessi,
 Int. J. Mod. Phys. {\bf A 10}, 1969 (1995).
 
\bibitem{BChanM2} A.P. Balachandran, L. Chandar and A. Momen,
\NP~{\bf B 461}, 581 (1996).

\bibitem{BBM2} A.P. Balachandran, G. Bimonte and G. Marmo,
\NP~{\bf B 446}, 299 (1995).

\bibitem{Asoreybal} M. Asorey, A.P. Balachandran and J.M. P\'erez-Pardo,
\JHEP {\bf 12}, 073 (2013).

\bibitem{Asoreyach} N. Acharyya, M. Asorey, A.P. Balachandran 
and S. Vaidya, \PR~{\bf D 92}, 105016 (2015).

\bibitem{skygroup} A.P. Balachandran and S. Vaidya,  Eur. Phys. J. Plus {\bf 128}, 118 (2013); A.P. Balachandran, S. {K\"{u}rk\c{c}\"{u}o\v{g}lu}, A.R. de Queiroz and
S. Vaidya, Eur. Phys. J. {\bf C 75}, 2, 89 (2015);
A.P. Balachandran and A.F. Reyes-Lega, in
Springer Proc. Phys. 229 (2019) 41-55.

\bibitem{BNV1} A.P. Balachandran, V.P. Nair and S. Vaidya,
\PR~{\bf D 100}, 045001 (2019).

\bibitem{edge-dual} A.P. Balachandran, V.P. Nair, A. Pinzul,
A.F. Reyes-Lega and S. Vaidya, \PR~{\bf D 106}, 025001 (2022).

\bibitem{asorey2} M. Asorey, A.P. Balachandran, M.A. Momen and
B. Qureshi, \JHEP~{\bf 10}, 028 (2023) [arXiv:2304.11008[hep-ph]].

\bibitem{edge-NCCS} A.P. Balachandran, K.S. Gupta and
S. {K\"{u}rk\c{c}\"{u}o\v{g}lu}, \JHEP {\bf 9}, 007 (2003).

\bibitem{string} See, for example, O. Aharony, S. Gubser, J. Maldacena, H. Ooguri and Y. Oz,
Phys. Rep. {\bf 323}, 183 (2000);
J. Polchinski, arXiv:1010.6134[hep-th].

\bibitem{loopg} See, for example, A. Ashtekar, 
Rep. Prog. Phys. {\bf 84}, 042001 (2021).

\bibitem{ncgeom} See, for example,
A. Connes, {\it Noncommutative Geometry}, Academic Press, Boston,
1994;
G. Landi, {\it An introduction to noncommutative spaces and their geometries}, Lecture Notes in Physics. New Series m: Monographs, vol. 51, Berlin, New York: Springer-Verlag [arXiv:hep-th/9701078].

\bibitem{causal} L. Bombelli, J. Lee, D. Meyer and R.D. Sorkin,
\PRL~{\bf 59}, 521 (1987);
R.D. Sorkin, Lectures at the School on Quantum Gravity,
Valdivia, Chile, January 2002 [arXiv:gr-qc/0309009].

\bibitem{entropicg} T. Jacobson, \PRL~{\bf 75}, 1260 (1995);
T. Padmanabhan, Rep. Prog. Phys. {\bf 73}(4), 6901 (2010);
E. Verlinde, \JHEP {\bf 2011} (4) 29 (2011).

\bibitem{twist-P} M. Dimitrijevic and J. Wess, {\it Deformed bialgebra of diﬀeomorphisms},
arXiv:hep-th/0411224; P. Aschieri, C. Blohmann, M. Dimitrijevic, F. Meyer,
P. Schupp and J. Wess, {\it A gravity theory on noncommutative spaces},
arXiv:hep-th/0504183.\\
M. Chaichian, P. P. Kulish, K. Nishijima and A. Tureanu, 
Phys. Lett. B 604, 98 (2004) [arXiv:hep-th/0408069]; M. Chaichian, P. Presnajder and A. Tureanu, 
Phys. Rev. Lett. {\bf 94}, 151602 (2005)
[arXiv:hep-th/0409096].\\
Robert Oeckl, 
Commun. Math. Phys. {\bf 217} (2001) 451-473 [arXiv:hep-th/9906225];
Nucl.Phys. B581 (2000) 559-574 [arXiv:hep-th/0003018].\\
H. Grosse, J. Madore, H. Steinacker,
J. Geom. Phys. {\bf 38}, 308 (2001);
J. Geom. Phys. {\bf 43}, 205 (2002).

\bibitem{BGMPQV} A.P. Balachandran, T.R. Govindarajan, 
G. Mangano, A. Pinzul, B. Qureshi and S. Vaidya,
\PR~{\bf D 75}, 045009 (2007).

\bibitem{BPJ} A.P. Balachandran, P. Padmanabhan and A. Joseph,
\PRL~{\bf 105}, 051601 (2010).

\bibitem{BGMT-S} A.P. Balachandran, T.R. Govindarajan, 
C. Molina and P. Teotonio-Sobrinho,
\JHEP {\bf 10}, 072 (2004).

\bibitem{B-Ib-MM} A.P. Balachandran and M. Martone,
Mod. Phys. Lett. {\bf A 24}, 1721 (2009);
A.P. Balachandran, A. Ibort, G. Marmo and M. Martone,
\PR~{\bf D 81}, 085017 (2010);
SIGMA {\bf 6}, 052 (2010);
\JHEP~1103:057 (2011).

\bibitem{UV-IR} S. Minwalla, M. van Raamsdonk and N. Seiberg,
\JHEP {\bf 0002}, 020 (2000);
for an updated analysis, see N. Craig and S. Koren,
\JHEP {\bf 03}, 037 (2020) and references therein.

\bibitem{BPQur} A.P. Balachandran,  A. Pinzul and B. Qureshi,
\PL~{\bf B 634}, 434 (2006).

\bibitem{NielN} H.B. Nielsen and M. Ninomiya, \PL~{\bf B 105}, 219 (1981);
\NP~{\bf B 185}, 20 (1981).

\bibitem{GKP}  H. Grosse, C. Klim{\v c}{\'i}k, and P. Pre{\v s}najder, \CMP~{\bf 178},
507 (1996); {\bf 185}:155 (1997); H. Grosse and P. Pre{\v s}najder,
Lett. Math. Phys. {\bf 46}, 61 (1998);
H. Grosse, C. Klim{\v c}{\'i}k, and P. Pre{\v s}najder, \CMP~{\bf 180},
429 (1996); H. Grosse, C. Klim{\v c}{\'i}k, and P. Pre{\v s}najder, in Les Houches Summer
School on Theoretical Physics, 1995, arXiv:hep-th/9603071. 

\bibitem{BGY1} A.P. Balachandran, T.R. Govindarajan and B. Ydri,
Mod. Phys. Lett. {\bf A 15}, 1279 (2000);
arXiv:hep-th/0006216.

\bibitem{BImm} A.P. Balachandran and G. Immirzi,
\PR~{\bf D68}, 065023 (2003).

\bibitem{BBVY} S. Baez, A.P. Balachandran, S. Vaidya and B. Ydri,
\CMP~{\bf 208}, 787 (2000);
A.P. Balachandran and S. Vaidya, Int. J. Mod. Phys.
{\bf A 16}, 17 (2001).

\bibitem{Klim} C. Klim{\v c}{\'i}k, \CMP~{\bf 199}, 257 (1998).


\bibitem{CP2F} G. Alexanian, A.P. Balachandran, G. Immirzi and B. Ydri,
J. Geom. Phys. {\bf 42}, 28 (2002).

\bibitem{CPNF} A.P. Balachandran, B. Dolan, J-H Lee, X. Martin and D. O'Connor, J. Geom. Phys. {\bf 43}, 184 (2002).

\bibitem{fuzzysusy} A.P. Balachandran, S. {K\"{u}rk\c{c}\"{u}o\v{g}lu} and E. Rojas,
\JHEP {\bf 07}, 056 (2002).

\bibitem{others} V.P. Nair and S. Randjbar-Daemi, \NP~{\bf B 533}, 333 (1998);
H. Grosse and A. Strohmaier, Lett. Math. Phys. {\bf 48}, 163 (1998).
For a construction in the context of the quantum Hall effect, see
D. Karabali and V.P. Nair, \NP~{\bf B 641}, 533 (2001);
{\bf B 679}, 427 (2004); J. Phys. {\bf A 39}, 12735 (2006).

\bibitem{contlim} A.P. Balachandran, X. Martin and D. O'Connor, 
Int. J. Mod. Phys. {\bf A 16}, 2577 (2001).

\bibitem{BSec} A.P. Balachandran and S. {K\"{u}rk\c{c}\"{u}o\v{g}lu},
Int. J. Mod. Phys. {\bf A 19}, 3395 (2004).

\bibitem{poset-sor} R.D. Sorkin, Int. J. Theor. Phys. {\bf 30}, 923 (1991).

\bibitem{poset-bal} A.P. Balachandran, G. Bimonte, E. Ercolessi, G. Landi,
F. Lizzi, G. Sparano and P. Teotonio-Sobrinho,
\NP~ Proc. Suppl. {\bf 37 C}, 20 (1995);
J. Geom. Phys. {\bf 18}, 163 (1996).

\bibitem{BQueir1} A.P. Balachandran and A.R. de Queiroz,
\JHEP ~{\bf 11}, 126 (2011);
\PR~{\bf D 85}, 025017 (2012).

\bibitem{BGQueir} A.P. Balachandran, T.R. Govindarajan and
A.R. de Queiroz,  Eur. Phys. J. Plus {\bf 127}, 118 (2012).

\bibitem{BQueirV} A.P. Balachandran, A.R. de Queiroz and S. Vaidya,
Eur. Phys. J. Plus {\bf 128}, 112 (2013);
\PR~{\bf D 88}, 025001 (2013).

\bibitem{BGQueirReyes} A.P. Balachandran, T.R. Govindarajan,
A.R. de Queiroz and A.F. Reyes-Lega, \PR~{\bf A 88}, 022301 (2013);
\PRL~{\bf 110}, 080503 (2013).

\bibitem{isospinballs} A.P. Balachandran, B. Rai and A.M. Srivastava,
\PRL~{\bf 59}, 853 (1987);
A.P. Balachandran, B. Rai, G. Sparano and A.M. Srivastava,
Int. J. Mod. Phys. {\bf A 3}, 2621 (1988).

\bibitem{ABJJQu} E. Akofor, A.P. Balachandran, S. Jo, A. Joseph and
B. Qureshi, \JHEP {\bf 05}, 092 (2008);
E. Akofor, A.P. Balachandran,  A. Joseph, L. Pekowsky and
B. Qureshi, \PR~{\bf D 79}, 063004 (2009).

\bibitem{BQueirV2} A.P. Balachandran, A.R. de Queiroz and S. Vaidya,
Mod. Phys. Lett. {\bf A 30}, 1550080 (2015);
Int. J. Mod. Phys. {\bf A 30}, 1550064 (2015).

\bibitem{Ach} N. Acharyya, A.P. Balachandran, 
M. Pandey, S. Sanyal and S. Vaidya, Int. J. Mod. Phys. {\bf A 33}, 1850073
(2018); N. Acharyya and A.P. Balachandran, \PR~{\bf D 96}, 074024 (2017);
see also M. Pandey and S. Vaidya, \PR~{\bf D 101}, 114020 (2020).

\bibitem{BDMat} A.P. Balachandran, S. Digal and T. Matsuura,
\PR~{\bf D 73}, 074009 (2006).

\bibitem{festschrift} {\it Particles, Fields and Topology: Celebrating
A P Balachandran}, T.R. Govindarajan, G. Marmo,
V. Parameswaran Nair, D. O'Connor, S.G. Rajeev and S. Vaidya (eds.),
World Scientific Publishing Co. Pte. Ltd., 2023.

\bibitem{fedele} Fedele Lizzi, in \cite{festschrift}, pp.117-120,
arXiv:2211.05532[physics.hist-ph].

\bibitem{bk-BMSS1} A. P. Balachandran, G. Marmo, B-S. Skagerstam, A. Stern, {\it Gauge Symmetries and Fibre Bundles- Applications to Particle
Dynamics}, Lecture Notes in Physics 188, Springer Verlag, Berlin 1983.

\bibitem{bk-BMSS2} A. P. Balachandran, G. Marmo, B-S. Skagerstam, A. Stern, {\it Classical Topology and Quantum States}, World Scientific Publishing Co. Pte. Ltd., Singapore, 1991.

\bibitem{bk-BKV} A. P. Balachandran, S. {K\"{u}rk\c{c}\"{u}o\v{g}lu}, S. Vaidya, {\it Lectures on Fuzzy and Fuzzy Susy Physics}, World Scientific Publishing Co. Pte. Ltd., Singapore, 2007. 

\bibitem{BalJM} A.P. Balachandran, S. Jo and G. Marmo,
{\it Group Theory and Hopf Algebras}, 
World Scientific Publishing Co. Pte. Ltd., Singapore,  2010.

\bibitem{BalT-group} A.P. Balachandran and C.G. Trahern,
{\it Lectures on Group Theory for Physicists}, Brill
Academic Publishing, 1986.

\end{thebibliography}
\end{document}